# Digital twins' kinetics of virtual free-radical copolymerization of vinyl monomers with stable radicals. 3. *N*-isopropyl acrylamide

Elena F. Sheka

Institute of Physical Researches and Technology, Peoples' Friendship University of Russia (RUDN University), 117198 Moscow, Russia;

sheka@icp.ac.ru

**Abstract:** The first experience of virtual free-radical copolymerization (FRCP) of vinyl monomers with stable radicals in the framework of the digital twins (DTs) concept (arXiv:2309.11616 [physics.chem-ph] and arXiv:2311.02752 [cond-mat.mtrl-sci]) is extended to *N*-isopropyl acrylamide (NIPA). The virtualization of the chemical process is based on the concept of a chain reaction that covers a set of elementary reactions (ERs) and is the most suitable for quantum chemical treatment. The calculations were performed using a semi-empirical version of the unrestricted two-determinant Hartree-Fock approximation. Once input in the reaction solution, providing free-radical polymerization of NIPA, the fullerene $C_{60}$ captures the formed monomer-radicals and terminates the polymerization. The obtained virtual kinetic data, providing the appearance of an induction period at the initial stage of the NIPA polymerization, are in full agreement with experimental reality.

**Keywords:** digitalization of polymerization; vinyl monomers; free-radical polymerization; digital twins approach; energy graphs; activation energy; *N*-isopropyl acrylamide; fullerene $C_{60}$

**1. Introduction**

The digitalization, recently proposed for the free-radical polymerization of vinyl monomers [1-3], rests on three pillars. First in this group is the concept of polymerization as a chain reaction consisting of a set of elementary reactions occurring in superposition [4-8]. This concept underlies a set of digital twins (DTs) [9,10] that represent the starting chemicals of the reactions and their final products. The second pillar concerns the consideration of a sufficiently large number of the reactions and their products in the same way, taking into account the spin features of the considered DTs, radical most of them. In the field of quantum chemical computing, tools based on the semiempirical two-determinant Hartree-Fock approximation has so far been best handled this role. A special format for working with a large set of results obtained, presented in the form of matrix tabulating, is the third pillar. This format allows us to propose an universal "polymerization passport" (PP) of the polymerizable medium under study. The availability and high facility of computational methods makes it possible to issue such a passport to almost any reaction system. The passports themselves, on the one hand, are purely individual with respect to any of the reaction participants. On the other hand, they represent an informative source of knowledge concerning the process under study. The first application of the PP approach took place for the free-radical polymerization (FRP) of methyl methacrylate and its free-radical copolymerization (FRCP) with fullerene C$_{60}$ and another stable radical TEMPO [2]. Then more detailed study was performed for styrene [3]. This work is devoted to the PP issue to the FRP of *N*-isopropyl acrylamide (NIPA) and its FRCP with C$_{60}$.

## 2. Polymerization passport of NIPA

Necessary information regarding the FRP and FRCP of vinyl monomers, the need to introduce PP and its advantages in describing specific systems have been discussed earlier [1-3]. In this work, the main attention is focused on issuing the PP for the polymerizable medium, the main monomer participant of which is NIPA. As mentioned, PP is an extremely individual "purely personal" document containing data related to a clearly defined set of chemical reagents in the reaction solution. In this work, we will talk about the NIPA FRP initiated by the free radical, $AIBN^{\bullet}$ produced in the course of the thermal decomposition of 2,2'-azobisisobutyronitrile (*AIBN*), and its FRCP with the C$_{60}$ fullerene. The data listed in Table 1 are the main part of the PP. The nomination of elementary reactions and/or corresponding DTs is disclosed in Table 2. The second mandatory page of the passport is given by Figure 1, which presents graphic images of equilibrium structures, the data for which are given in Table 1. As seen, this table consists of three parts. The first parts contains the above mentioned nominations, related to both reactions and the relevant DTs that are important for the consideration of the FRP and FRCP under study. The second part involves thermodynamic descriptors related to the considered reactions expressed in terms of coupling energies $E_{cpl}$ of their final products. The third part presents kinetic descriptors of the studied reactions expressed in terms of association activation energies $E_{aa} \equiv E_a$ [1]. All the digit data are obtained exploiting software CLUSTER-Z1 [11,12], based on the semiempirical AM1 version of the unrestricted two-determinant Hartree-Fock (UHF) approximation [13]. Bold-marked $E_a$ values listed in the table were evaluated when constructing the relevant barrier profiles of the DTs decomposition (see detailed description of the technique elsewhere [1-3]).

Figure 1 presents equilibrium structures of the considered DTs related to the FRCP of NIPA with C$_{60}$. The main participants of the reaction, which are the headings of Table 1 added with $M$, are given in Figure 1a. NIPA oligomer-radicals, concerning the FRP of the latter (data in the yellow cells of the table), are given in [1] and are not presented in the figure. In contrast, all four members of the NIPA FRCP with C$_{60}$ (the content of light pink cells of the table) are given in Figure 1b. The barrier profiles of the dissociation of the latters are shown in Figure 2a and b. Among the latters, the attention should be particularly given to Figure 2a. It exhibits results of the decomposition of fullerenyl $FM$, which is formed by a two-dentant coupling of NIPA molecule with fullerene. The corresponding intermolecular junction is of 2,2-cyclo addition form. Evidently, the junction breaking concerns two *sp³*C-C bonds. This can happen by elongation and breaking of both bonds either simultaneously, or one-by-one sequentially. These two ways are presented in the figure revealing a quite significant difference in the $E_a$ values in the two cases. At the same time, as seen in the figure, the maximum positions of all the graphs are clearly vivid, thus evidencing the position of the transition state $E_{ST}$ inside the region, marked with blue band. The latter marks the dispersion of the $R_{crit}^{C-C}$ of 2.11±0.1 Å that determines maximum length of the *sp³*C-C bond, above which the bond becomes radicalized thus revealing the start of its breaking [14,15]. Figure 2b presents the barrier profiles of the decomposition of three monofullerenyls $FM^{\bullet}, FR^AM$, and $FR^A$, which alongside with the above considered $FM$ constitute a fullerene C$_{60}$ family of DTs, once of interest from the standpoint of the reaction under consideration. As can be seen from the figure, in all the cases, the DT decomposition is caused by the rupture of a single

$sp^3$C-C bond, accompanied by a significant dispersion of the critical value of its length. This somewhat blurs the structure of the graphs under study in the region of their maxima and makes the determination of the critical value of the bond length less certain. Noteworthy is the fact that the $FR^AM$ dissociation turns out to be barrier-free.

**Table 1.** Elementary reactions and their final DT products supplemented with virtual thermodynamic and kinetic descriptors related to the FRCP of NIPA with fullerene C$_{60}$, while initiated with $AIBN^\bullet$ free radical

|  | $R^AM^\bullet$ | $AIBN^\bullet$ $R^{A\bullet}$ | $C_{60}$ $F$ | |
|---|---|---|---|---|
| **Digital twins** | | | | |
|  |  |  | 2-dentate | 1-dentate |
| $M$ | $R^AM_2^{\bullet\bullet}$ | $R^AM^\bullet$ | $FM$ | $FM^\bullet$ |
| $R^AM^\bullet$ | $(R^AM)_2$ | $R^AR^AM$ | $FR^AM$ | |
| $R^{A\bullet}$ | - | - | $FR^A$ | |
| **Thermodynamic descriptors $E_{cpl}$, kcal/mol** [1] | | | | |
| $M$ | -25,378 (2) -23.51 (3) -24.05 (4) | −5.675 (1) | -26.483 | 0.598 |
| $R^AM^\bullet$ | - | - | -40.161 | |
| $R^{A\bullet}$ | - | - | -20.137 | |
| **Kinetic descriptors $E_a$, kcal/mol** [1,2] | | | | |
| $M$ | **8.39 (2)** 7.74 (3) [3] 8.49 (4) [3] | **19.09 (1)** | **17.29 (1)** [4] **27.79 (2)** | **20.01** |
| $R^AM^\bullet$ | - | - | **0.023** | |
| $R^{A\bullet}$ | - | - | **9.398** | |

[1] Digits in brackets mark the number of monomers in the oligomer chain.
[2] Bold data are determined from the barrier profiles similar to shown in Figure 2.
[3] Non-bold data are calculated by using Evans-Polanyi-Semenov relation presented in [3].
[4] Digits in brackets indicate one-dentant and two-dentant coupling of NIPA with C$_{60}$, respectively.

## 3. Virtual and real kinetics of the FRCP of NIPA with fullerene C$_{60}$

NIPA stands somewhat apart from the bulk of vinyl monomers, which is due to the water solubility of its copolymers with C$_{60}$ fullerene, in contrast to the dominant majority of vinyl polymers that are insoluble in water. Water-solubility is of particular interest due to the promise of NIPA polymers use in biology and medicine, for example, as nanocontainers for the delivery of drugs to 'sick' cells. They can also be used for water purification from microimpurities or extraction from aqueous solutions of valuable substances present in low concentrations. The copolymerization reaction of NIPA with fullerene C$_{60}$ is described in detail in works [17,18]. In this study, special attention is paid to the kinetics of the initial stage of the reaction, represented in terms of the time dependence of the percentage monomer conversion $x(t)$ in Figure 2c. As seen in the figure, the evident feature of the dependence concerns the long IP of practically zero amplitude with respect to the abscissa. The IP length is the largest of known for vinyl polymers up today. It means that input of small additive of $C_{60}$ into the chemicl reactor terminates the NIPA polymerization for about several hours, after which a rapid occurrence of the process takes place similarly to that observed for styrene or methyl methacrilate, for example.

**Table 2.** Elementary reactions of the initial stage of the free-radical copolymerization of NIPA with fullerene $C_{60}$

| Reaction mark | Reaction equation [1] | Reaction rate constant | Reaction type |
|---|---|---|---|
| (1) | $R^\bullet + M \to RM^\bullet$ | $k_i$ | generation of monomer-radicals |
| (2) | $RM^\bullet + (n-1)M \to RM_n^\bullet$ | $k_p$ | generation of oligomer-radicals, polymer chain growth |
| (3a) | $F + M \to FM$ | $k_{2m}^F$ | two-dentant grafting of monomer on $C_{60}$ |
| (3b) | $F + M \to FM^\bullet$ | $k_{1m}^F$ | one-dentant stable radical grafting of monomer, generation of monomer-radical |
| (4) | $FM^\bullet + (n-1)M \to FM_n^\bullet$ | $k_p^F$ | generation of oligomer-radical anchored to $C_{60}$, polymer chain growth |
| (5) | $F + RM^\bullet \to FRM$ | $k_{rm}^F$ | monomer-radical grafting on $C_{60}$ |
| (6) | $F + R^\bullet \to FR$ | $k_R^F$ | free radical grafting on $C_{60}$ |
| (7) | $R^\bullet + FM^\bullet \to RFM$ | $k_{FM}^R$ | monomer-radical $FM^\bullet$ capturing with free radical |

[1] $M, R, F$ mark DT's chemical attribution to NIPA, initiating free radical $AIBN^\bullet$, and fullerene $C_{60}$, respectively. Superscript black spot distinguishes radical species.

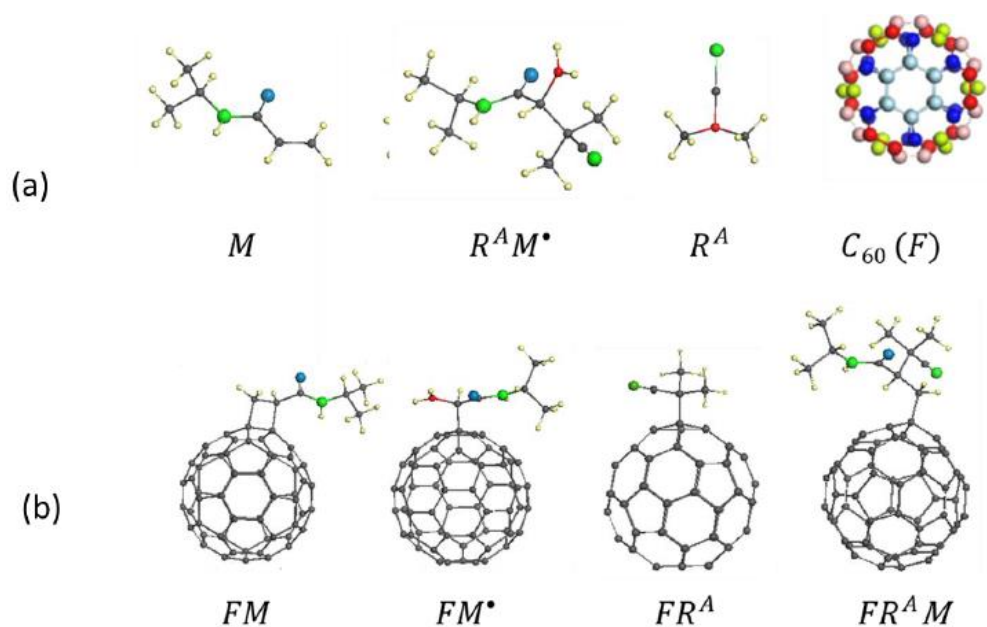

(a) M, $R^A M^\bullet$, $R^A$, $C_{60}$ (F)

(b) FM, $FM^\bullet$, $FR^A$, $FR^A M$

**Figure 1.** Equilibrium structures of digital twins related to the FRCP of NIPA with fullerene $C_{60}$. (a) Digital twins, representing headings of Table 1. Colored image of $C_{60}$ exhibits a specific spin density distribution over the molecule carbon atoms, thus marking their variable chemical activity [16]. (b) Digital-twin monofullerenyls produced by intermolecular interaction of $C_{60}$ with NIPA, both two-dentant $FM$ and one-dentant $FM^\bullet$ coupling as well as by trapping of either free radical ($AIBN^\bullet$) $FR^A$ or monomer-radical $FR^A M$. UHF AM1 calculations.

Considering the picture of the kinetics of FRCP, presented by the values of $E_a$ in Table 1, one can clearly conclude that it is this type of reaction, similar to that observed experimentally, that could be expected for NIPA in the presence of small additions of fullerene. Indeed, the active participation of fullerene in copolymerization is determined by three elementary reactions $FM^\bullet$, $FR^A$, and $FR^AM$. The first one determines the polymerization of NIPA on fullerene, caused by the formation of a monomer-radical $FM^\bullet$. As can be seen from the table, the formation of this fullerenyl is accompanied by a positive coupling energy, as a result of which $E_{ad} > E_{aa}$, so that the generation of $FM^\bullet$ is kinetically unfavorable. In contrast to the case, the formation of fullerenyls $FR^A$, and $FR^AM$ is kinetically favorable, but the huge difference between their activation energies makes it possible to simply forget about the first of them. Thus, in a reaction solution, containing fullerene C$_{60}$, the $FR^AM$ reaction has an evident kinetic advantage, which determines the capture of the $R^AM^\bullet$ monomer-radical by the fullerene and thus stops the polymerization of the monomer. The extremely low value of its activation energy makes this capture practically barrier-free, while the following chain of the rate constants related to the fullerene-associated reactions looks like

$$k_{rm}^F \gg k_R^F \gg k_{1m}^F.$$

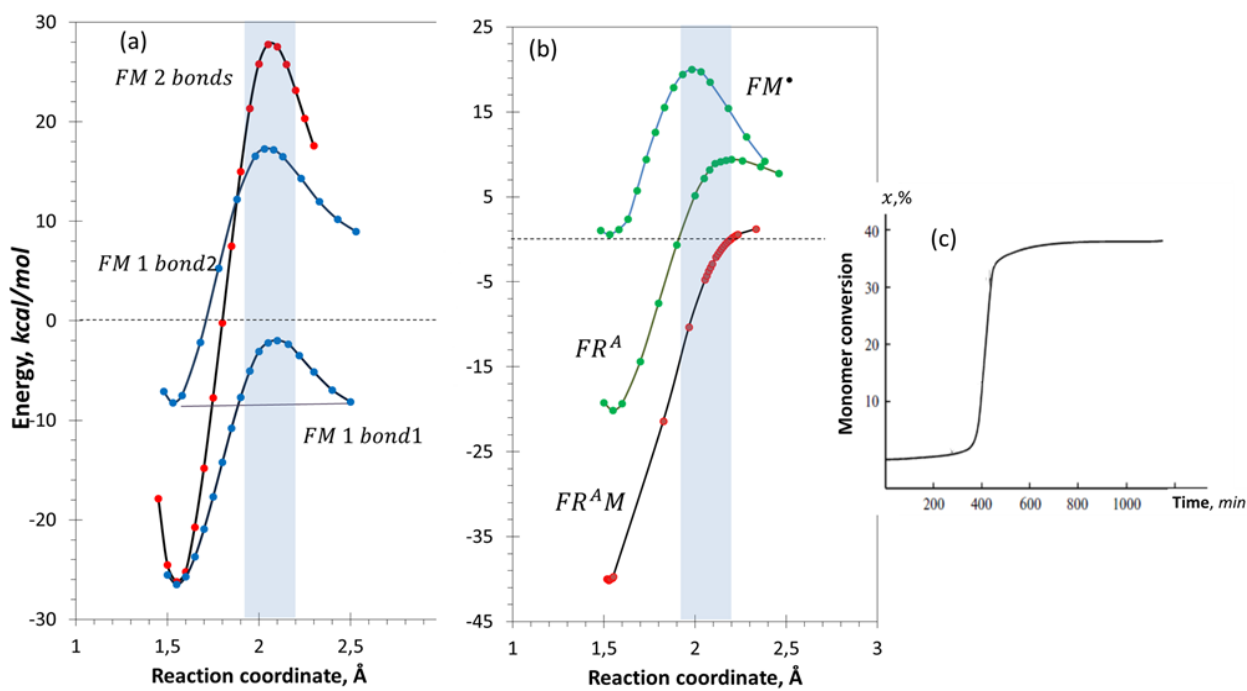

**Figure 2.** (a) and (b) Energy graphs, presented virtual barrier profiles of the decomposition of the following DTs: monofullerenyl $FM$, related to two-dentant coupling of NIPA with fullerene and decomposed either through elongation of the two bonds simultaneously, or through one-by-one bond sequentially (a); monofullerenyls $FM^\bullet, FR^A$ and $FR^AM$ related to the reactions of the same name. UHF AM1 calculations. (c) Empirical kinetics of the $AIBN^\bullet$-initiated FRCP of NIPA with fullerene C$_{60}$. T=60$^0$C; o-DCB solvent; [NIPA]=0.73 mol/L; [AIBN]=0.24 mol/L; [C$_{60}$]=6.7x10$^{-3}$mol/L. Digitalized data of Ref. 17.

This is in perfect agreement with the reality: the addition of fullerene stops the free-radical polymerization of NIPA, which resumes after the added C$_{60}$ is fully consumed.

## 4. Conclusion

This work completes a series of studies of virtual free-radical copolymerization of vinyl monomers with small additives of stable radicals, such as *TEMPO* and $C_{60}$ fullerene [2,3]. The digitalization of the thermodynamic and kinetic descriptors of this process, undertaken for the first time, showed that these radicals influence free-radical polymerization of the bodies completely differently. The input of *TEMPO* inhibits the latter in all the cases manifesting itself in the same way, regardless of the chemical composition of the reaction solutions. The inhibition capacity is dictated by a single elementary reaction of capturing the monomer-radical $RM^{\bullet}$ initiated by a free one, the rate constant of which ($k_{rm}^S$) occurs the highest. In practice, this feature is revealed by the presence of an induction period in the time dependence of monomer conversion $x(t)$ in the initial stage of polymerization with the amplitude of its change close to zero, while its duration is governed with the TEMPO content [19,20].

In the case of fullerene $C_{60}$, the situation is drastically changed. Not one, but three elementary reactions may influence the FRP of monomers. These are the formation of monomer-radical $FM^{\bullet}$, initiated with fullerene ($k_{1m}^F$); the capture of free radicals-initiators $R^{\bullet}$ with $C_{60}$ $FR$ ($k_R^F$); and the capture of monomer-radical $RM^{\bullet}$ with fullerene $FRM$ ($k_{rm}^F$). All the reactions provide the formation of different fullerenyls, thermodynamics and kinetics of which depends on the chemicals attached to the $C_{60}$ core according to the general rules of the fullerene nanochemistry [16]. Actually, the magnitudes of the above rate constants respond to changing in chemical content right away, which causes a change in the arrangement of each of these constants in their order from large to small. Thus, the rate constant $k_R^F$ leads the order in the case of methyl methacrylate, polymerized being initiated with $AIBN^{\bullet}$; $k_{1m}^F$ does the same in the case of styrene, but is replaced with $k_{rm}^F$ when $AIBN$ is replaced with benzoyl peroxide; $k_{rm}^F$ determines first steps of the FRP of NIPA, initiated with $AIBN^{\bullet}$. Evidently, the FRCP of vinyl monomers with fullerene $C_{60}$ cannot be described with a unique appearance in all the cases, while the latter is highly individual depending on atomic structure of monomers and free radicals, initiating the reaction.

The virtual variability of the first stage of the FRCP of the monomers with fullerene $C_{60}$ discussed above finds its confirmation in reality. Thus, the domination of the reaction $FR$ means the reduction of the current concentration of free radicals, which should slow down the polymerization of monomers and should be represented empirically as a decrease in the slope of the $x(t)$ curve with respect to the time axis. The decrease should progress when the concentration of $C_{60}$ increases [2]. It is this behavior of the $x(t)$ dependencies that are observed empirically for the FRCP of methyl methacrylate, initiated with $AIBN^{\bullet}$, in the presence of $C_{60}$ [19].

The domination of $k_{1m}^F$ in the case of the FRCP of styrene with $C_{60}$ means that the first reaction, occurred in the chemical solution with the monomer, $AIBN$ and fullerene, is the formation of the monomer-radical $FM^{\bullet}$, which provides the polymerization of styrene on fullerene. The reaction proceeds until all the amount of fullerene is consumed, after which the routine styrene polymerization proceeds as if it were in the FRP case. Accordingly, $x(t)$ curve should contain two fractions related to two ways of the monomer polymerization, the former related to the polymerization stimulated with and occurred on fullerene and the latter once pretty similar to the FRP of styrene under the same conditions. In reality, the former fraction of the $x(t)$ curve should be of extremely low intensity because of three-order magnitude difference in the concentration of fullerene and monomer [3]. It is this type of the $x(t)$ curves that is obtained empirically for the FRCP of styrene, initiated with $AIBN^{\bullet}$, in the presence of $C_{60}$ [19].

The domination of the $FRM$ reaction characterized with rate constant $k_{rm}^F$ was revealed virtually twice, related to styrene [3] and NIPA. In the former case, as was said, it occurs when $AIBN$ initiator is substituted with benzoyl peroxide. In the NIPA case, $AIBN$ remains at place. Obviously, the occurrence of the $FRM$ reaction means a complete impossibility of the monomer polymerization, because of which $x(t)$ curve should begin with a classical induction period with close-to-zero amplitude with respect to the time axis. The duration of the period is governed with the fullerene concentration since the monomer polymerization can start only after all the $C_{60}$ content is consumed. This polymerization itself should be identical to that one occurred in the absence of fullerene. It is this type of $x(t)$ curves that are observed empirically for styrene, initiated with $BP^\bullet$ [20] and NIPA [17].

The current study showed that basic chain-reaction concept involving a discrete set of independent elementary reactions is reasonably valid for the FRP of vinyl monomers. Following from this "polymerization-passport"-based approach is highly productive, on the one hand, and allows to share friendly the responsibility of the polymerizable system study between virtual and real experiments. Modern quantum chemistry is able to construct required passports within wide limits. As a consequence, digitalization of polymerization reactions may prove to be a broad platform for introducing this approximation into polymer science and technology.

**Acknowledgments.** This paper has been supported by the RUDN University Strategic Academic Leadership Program.

**References**

1. Sheka, E.F. Virtual free radical polymerization of vinyl monomers in view of digital twins. *Polymers* 2023, **15**, 2999.
2. Sheka, E.F. Digital twins kinetics of virtual free-radical copolymerization of vinyl monomers with stable radicals. 1. Methyl methacrylate. *arXiv*:2309.11616 [physics.chem-ph], 2023, https://doi.org/10.48550/arXiv.2309.11616.
3. Sheka, E.F. Digital twins' kinetics of virtual free-radical copolymerization of vinyl monomers with stable radicals. 2. Styrene. arXiv:2311.02752 [cond-mat.mtrl-sci]. https://doi.org/10.48550/arXiv.2311.02752
4. H.W. Starkweather, H.W.; Taylor, G.B. The kinetics of the polymerization of vinyl acetate. *JACS* 1930, **52**, 4708-4714.
5. Semenov, N.N. *Tsepnyie Reakcii* (Chain Reactions) Goschimizdat: Moskva, 1934 (in Russian).
6. Bagdasar'yan, Kh.S., *Teoriya radikal'noi polimerizatsii* (Free_Radical Polymerization Theory), Moscow: Nauka, 1966. (in Russian).
7. Gol'dfein, M.D.; Kozhevnikov, N.V.; Trubnikov, A.V. *Kinetika i mekhanizm regulirovaniya protsessov obrazovaniya polimerov* (Kinetics and Control of Polymerization Processes), Saratov: Saratov. Gos. Univ., 1989.
8. Pross, A. *Theoretical and Physical Principles of Organic Reactivity*, Wiley, New York, 1995.
9. Rasheed, A.; San, O.; Kvamsdal,T. Digital twins: Values, challenges and enablers from a modeling perspective. *IEEE Access* 2020, doi: 0.1109/ACCESS.2020.2970143.
10. Sheka, E.F. Digital Twins in graphene's technology. arXiv.2208.14926.
11. Zayets, V. A. *CLUSTER-Z1: Quantum-Chemical Software for Calculations in the s,p-Basis.* Inst. Surf. Chem. Nat. Ac. Sci. of Ukraine: Kiev, **1990**; (in Russian).

12. Berzigiyarov, P.K.; Zayets, V.A.; Ginzburg, I.Ya.; et al. NANOPACK: Parallel codes for semiempirical quantum chemical calculations of large systems in the *sp-* and *spd*-basis. *Int. J. Quantum Chem*. 2002**,** **88**, 449-462.
13. Dewar M.J.S.; Zoebisch, E.G.; Healey, E.F.; Stewart, J.J.P. AM1: A new general purpose quantum mechanical molecular model. *J.Amer.Chem.Soc.* 1985, **107**, 3902-3909.
14. Sheka, E.F. Stretching and breaking of chemical bonds, correlation of electrons, and radical properties of covalent species, <u>Adv. Quant. Chem</u>. 2015, **70**, 111-161.
15. Sheka, E.F. Private Archive of Computed Data, 2016, partially published [46,49].
16. Sheka, E.F. *Fullerenes. Nanochemistry, Nanomagnetism, Nanomedicine, Nanophotonics*. CRC Press, Taylor and Francis Group, Boca Raton, 2011.
17. Atovmyan, E.G.; Grishchuk, A.A.; Estrina, G.A.; Estrin, Ya.I. Formation of star-like water-soluble polymeric structures in the process of radical polymerization of *N*-isopropylacrylamide in the presence of $C_{60}$. *Russ. Chem. Bull., Int. Ed*. 2016, **65***, 2082—2088.*
18. Atovmyan, E. G. On the relationship between the fullerene reactivity and degree of substitution. *Russ. Chem. Bull., Int. Ed*. 2017, **66***, 567-570.*
19. Yumagulova, R. Kh.; Kuznetsov, S. I.; Diniakhmetova, D. R.; Frizen, A. K.; Kraikin, V. A.; Kolesov. S. V. On the initial stage of the free-radical polymerizations of styrene and methyl methacryate in the presense of fullerene $C_{60}$. Kinetics and Catalysis 2016, **57**, 380–387.
20. Yumagulova, R. Kh.; Kolesov, S.V. Specific features of reactions between fullerene $C_{60}$ and radicals stabilized by conjugation in the process of radical polymerization. Bullet, Bashkir University 2020, **25**. 47-51. DOI: 10.33184/bulletin-bsu-2020.1.8